\documentclass[sigconf]{acmart}

\usepackage{algorithmic}
\usepackage{graphicx}
\usepackage{textcomp}
\usepackage{hyperref}
\usepackage{tikz}
\usepackage{xcolor}
\usepackage{tabularx}
\usepackage{multicol}
\usepackage{multirow}
\usepackage{booktabs} 
\usepackage{color}
\usepackage[utf8]{inputenc}
\usepackage{soul}

\newcommand{\fakeparagraph}[1]{\smallskip\noindent\textbf{#1.}}
\newcommand*\circled[1]{\tikz[baseline=(char.base)]{
            \node[shape=circle,draw,inner sep=1pt] (char) {#1};}}

\AtBeginDocument{%
  \providecommand\BibTeX{{%
    \normalfont B\kern-0.5em{\scshape i\kern-0.25em b}\kern-0.8em\TeX}}}

\begin{document}

\setcopyright{none}

\title{A Demonstration of Smart Doorbell Design Using Federated Deep Learning}

\author{Vatsal Patel, Sarth Kanani, Tapan Pathak}
\email{vatsal.pce18@sot.pdpu.ac.in}
\email{sarth.kce18@sot.pdpu.ac.in}
\email{tapan.pce18@sot.pdpu.ac.in}
\affiliation{%
  \institution{Pandit Deendayal Petroleum University}
  \city{Gandhinagar}
  \country{India}
}

\author{Pankesh Patel, Muhammad Intizar Ali,}
\author{John Breslin}
\email{pankesh.patel@insight-centre.org,}
\email{ali.intizar@nuigalway.ie, john.breslin@nuigalway.ie}
\affiliation{%
  \institution{Confirm SFI Research Centre for Smart Manufacturing, Data Science Institute, NUI Galway, Ireland}
}

\begin{abstract}
Smart doorbells have been playing an important role in protecting our modern homes. Existing approaches of sending video streams to a centralized server (or Cloud) for video analytics have been facing many challenges such as latency, bandwidth cost and more importantly users' privacy concerns. To address these challenges, this paper showcases the ability of an intelligent smart doorbell based on Federated Deep Learning, which can deploy and manage video analytics applications such as a smart doorbell across Edge and Cloud resources. This platform can scale, work with multiple devices, seamlessly manage online orchestration of the application components. The proposed framework is implemented using state-of-the-art technology. We implement the Federated Server using the Flask framework, containerized using Nginx and Gunicorn, which is deployed on AWS EC2 and AWS Serverless architecture. 
\end{abstract}



\keywords{Federated Learning, Internet of Things, Video Analytics, Artificial Intelligence, Deep Learning, Machine Learning, Privacy, Security}


\maketitle

\section{Introduction}\label{sec:intro}


The smart doorbell has been playing an important role in protecting our modern homes since they were invented. The recent trend from big companies~\cite{Delaney2020} is to offer a smart doorbell that integrates all possible services including face recognition at the door. A common approach, adopted by these offerings, is to send image streams over the network to a central server (or Cloud), where all the processing takes place and appropriate decisions are made. Although this approach reduces the maintenance cost by keeping the application logic in one central location, it may not be suitable for applications relying on video analytics. Some of the reasons are: \textbf{First}, the central server approach for video analytics may not be suitable for latency-sensitive applications because of the delay caused by transferring data to a central server for analysis and back to the application. \textbf{Second}, the use of the central sever for continuous data storage, object detection, and analysis is expensive because these applications generate high volume of image and video data. Furthermore, the processing and storage of multiple video streams make the subscription more costly. Secondly, this design requires a huge amount of reliable bandwidth, which may not always be had. \textbf{Third}, even if we assume that we could address latency and bandwidth issue by empowering a sophisticated infrastructure, a large class of video-based applications may not be suitable because of regulations and security concerns of sharing data as there is an involvement of biometric data of residents. For instance, GDPR restricts the sharing of users' private data across organizations.







The recent advancements in Federated Learning~\cite{federated-learning2016, hfl} have shown the potential to address the aforementioned challenges. Federated Learning works on \textbf{\textit{model aggregation rather than data aggregation}} principle~\cite{liu2020fedvision}. Building a model using Federated Learning fits the problem naturally for video analytics applications: \textbf{First}, it trains the model(s) locally and then uploads the model parameters to a centralized server for aggregation. Thus, it prevents data leakage as sensitive data does not leave the smart doorbell device. \textbf{Second}, it reduces communication cost~\cite{federated-learning2016, elgamal2020sieve, 10.1145/3213344.3213345}, as devices upload the trained model parameters to the centralized server, instead of the images. Federated Learning is not much tested in practice so far, specifically for video analytics applications~\cite{liu2020fedvision}, thus some open questions related to implementation details for video analytics applications~(such as a potential architecture when it is applied to computer vision applications and an implementation of this approach for resource constrained IoT devices) need to be addressed.

In this paper, we showcase the ability of an intelligent framework~\cite{7809306, 7460669} based on Federated Learning~(addressing the challenges as mentioned above), which  can deploy and manage video analytics applications such as a smart doorbell across Edge and Cloud resources~\cite{10.1145/3041021.3054736}.  The proposed framework is implemented using state-of-the-art technology. We implement the Federated Server using the Flask framework, containerized using Nginx and Gunicorn deployed on AWS EC2 and AWS Serverless architecture. Second, we have built  MobileNet object detection models~\cite{10.1145/3410992.3411013, 10.1145/3423423.3423465} for different scenarios~(such as face detection, an unsafe content detection, a noteworthy vehicle detection) and deployed them on resource-constrained IoT devices using TensorFlow Lite to reduce the object detection latency. These models are developed using Federated Learning, as a novel distributed deep learning approach, on a popular datasets such as ImageNet, Common Objects in Context~(COCO).

\begin{figure*}[h]
  \centering
  \includegraphics[height=5.5cm, width=\linewidth]{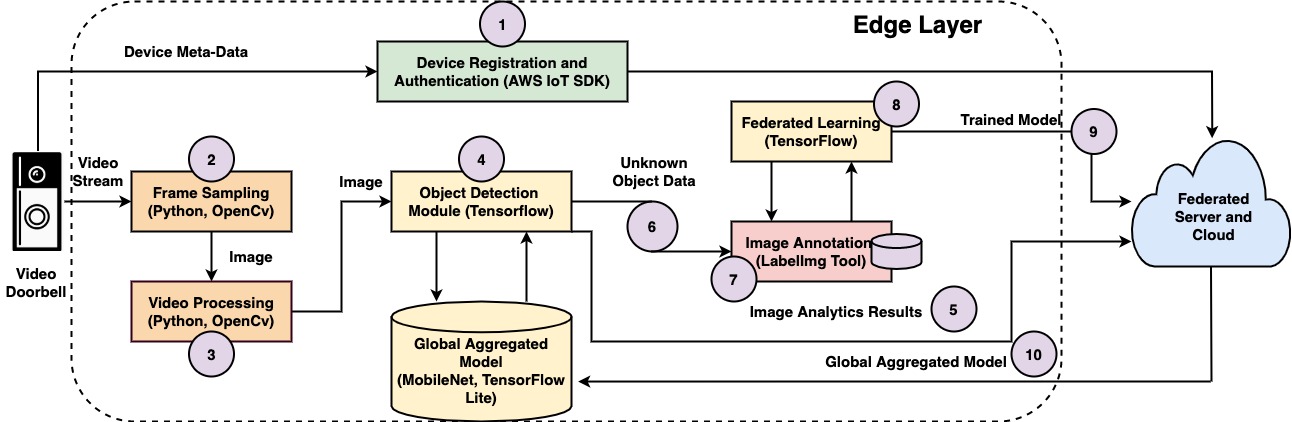}
  \caption{Logical Flow of Federated Learning for Video Analytics at Federated Client.}
  \label{fig:device-edge}
\end{figure*}

\section{System Design and Implementation}\label{sec:design}



The proposed system consists of \textbf{Federated Clients }and a \textbf{Federated Server}. The data flow goes as follows: A real-time video stream is captured by a camera and pre-processed at the Federated Client. It implements the video analytics logic to identify objects and training module to train a local model to be sent to the Federated Server. The Federated Server receives local models from each smart doorbell device and generates a global aggregated model. It distributes the aggregated global model back to the Federated Clients. The Federated Client uses this aggregated model to detect objects. The video analytics results from the Federated Client are sent to the Cloud layer for storage. This lets users access the doorbell anywhere and anytime. In the following, we present the functionality of each component and its implementation in detail.





\subsection{Federated Client}\label{sec:device-layer}
Each smart doorbell is interfaced with a camera module to capture a video stream and PIR sensor to detect the motion of an object. We prototype the smart doorbell using WiFi-enabled Raspberry Pi 3 Model B+. Each smart doorbell hosts the Federated Client. In the following section, we present the software components of the Federated Client.




\fakeparagraph{Device Registration and Authentication} Each Federated Client implements device registration and authentication, which allows users to interact with the device anywhere and anytime~(Circled~\circled{1} in Figure~\ref{fig:device-edge}) in a secure manner. We implement it using AWS IoT Core. The device registry keeps a record of all registered devices. Moreover, it supports X.509 certificate-based authentication so that data is never exchanged without proven identity~\cite{intizar:emse-01644333}.


\fakeparagraph{Frame Sampling}
It samples a frame off of a live video stream from the camera attached with the Federated Client~(Circled~\circled{2} in Figure~\ref{fig:device-edge}). It packages the captured frames and sends raw footage to the video pre-processing component for further pre-processing. 



\fakeparagraph{Video Pre-processing} A considerable part of a video stream contains data that is not useful. This consumes a huge chunk of a network's bandwidth and adds to computation cost unnecessarily. We employ spatial and temporal redundancy~\cite{FederatedVideoAnalytics} to remove redundant and uninteresting parts~(Circled~\circled{3} in Figure~\ref{fig:device-edge}):


\fakeparagraph{-- \textit{Temporal redundancy}} It reduces consecutive and similar video frames, using various filtering techniques such as motion detection. The motion sensor triggers the camera if there is any motion in front of the doorbell. The integration of a motion sensor allows the Federated Client to process data only when there is a motion.



\fakeparagraph{-- \textit{Spatial redundancy}} It is represented by removing the background of a video frame, which is not always necessary for object detection. We employ background subtraction technique~\cite{cv-rpi-book} to extract the Region of Interest~(RoI). It separates out foreground objects from the background. This technique is quite relevant for our smart doorbell as the background of an image largely remains uniform due to static camera.  The RoI is sent to the object detection module for further processing, as discussed in Section~\ref{sec:fl-edge}.

\begin{figure*}[h]
  \centering
  \includegraphics[height=8.1cm, width=\linewidth]{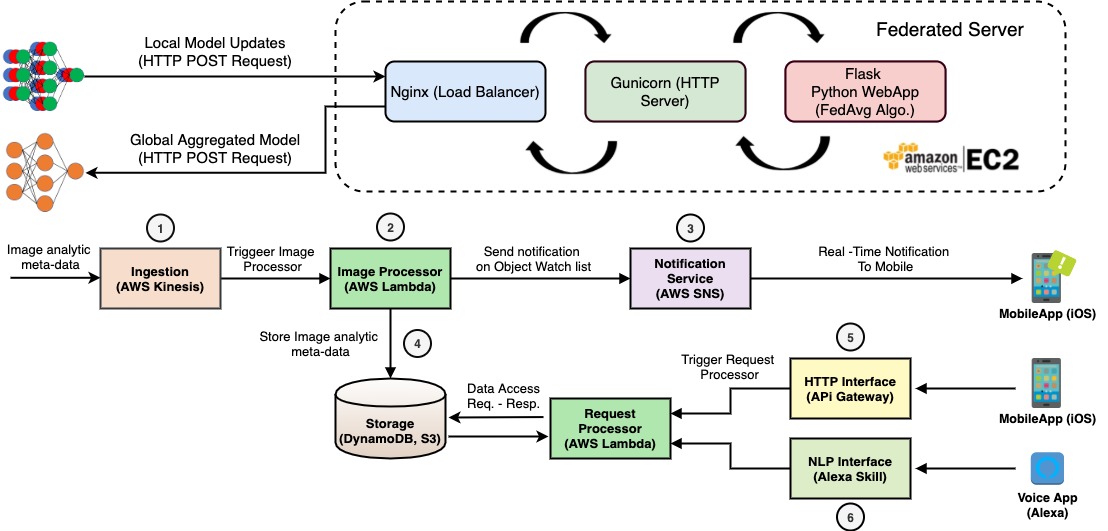}
  \caption{Logical Flow of Federated Server hosted on AWS EC2~(Upper part). AWS Serverless Architecture~(Lower Part).}
  \label{fig:fl-server-cloud}
\end{figure*}

\subsection{Federated Learning}\label{sec:fl-edge} 
This component runs the  \textit{Federated Learning} modules to train the object detection model locally, which are sent to the Federated Server for aggregation, and the \textit{Object Detection} module that uses an aggregated model from the Federated Server to detect objects.

\fakeparagraph{DL-based Object Detection} It is dedicated to running various object detection models. It takes the image as input from the video pre-processing module and runs various models to detect objects~(Circled~\circled{4} in Figure~\ref{fig:device-edge}). The current version implements four models: face detection and recognition, animal detection, unsafe content detection~(such as violence, gun etc.) and a noteworthy vehicle detection such as a fire truck and a courier service~(e.g., FedEx, USPS) van. For object detection, we adopt On-Device DL-approach. This approach employs various model reduction techniques~\cite{deeplearning-edge} (e.g., model compression, parameter pruning, parameter quantization, model design) to enable its deployment on IoT devices, while maintaining a reasonably good object detection accuracy. The current implementation uses MobileNets ~\cite{howard2017mobilenets}, which is a family of computer vision models for TensorFlow, designed for resource-constrained devices such as mobile phones and embedded devices. 


Depending on the detection results, the object detection module decides whether data needs to be sent to the Cloud layer or it is to be kept in local memory of the doorbell. For instance, if an object is identified by this module, the video analysis meta-data is sent to Cloud~(Circled~\circled{5} in Figure~\ref{fig:device-edge}). The image is stored in local memory in case the object is identified as new or unknown. The stored images are processed further by the training module~(Circled~\circled{6} in Figure~\ref{fig:device-edge}), as discussed in the next section.

\fakeparagraph{Federated Learning}
This component is responsible for two tasks: first, image annotations to label locally stored images; second, the Federated Learning module uses these annotated images to build local models, typically contains model parameters and corresponding weights~(Circled~\circled{9} in Figure~\ref{fig:device-edge}). The image annotation module~(Circled~\circled{7} in Figure~\ref{fig:device-edge}) provides an interface that lets the smart doorbell owners specify a bounding box and the corresponding label information, similar to the work~\cite{liu2020fedvision}. This image annotation process requires the smart doorbell user to be able to visually identify where the objects of interest are located in a given image file and draw the bounding box and assign it to a category. We integrate LabelImg tool~\cite{darrenl} to implement this functionality. This tool generates annotations as an XML file, which is automatically mapped to an appropriate system directory for model training.

\subsection{Federated Server at Cloud}\label{sec:federated-server}




It receives model updates learned at  Client. It performs model aggregation on them to produce a global aggregated model and distributes back it in the federation to be used for inference in object detection operations~(Circled~\circled{10} in Figure~\ref{fig:device-edge}). 

The model aggregation algorithm leverages Horizontal Federated Learning~(HFL)~\cite{hfl}. It can be applied in collaborative learning scenarios in which the device shares the same feature space but it is collected from different devices. HFL is suitable for our smart doorbell application scenario as it aims to help multiple devices with data from the same feature space (i.e., labelled image data) to train a global aggregated object detection model. The algorithm performs component-wise parameter averaging which are weighed based on the proportion of data points contributed by each participating doorbell device. The following is federated averaging equation~\cite{fedavgalgo}:

\begin{equation*}
f(w) = \sum\limits_{k=1}^{K}  \frac{n_k}{n} F_k(w) 
\quad \textrm{where} \quad
F_k(w) = \frac{1}{n_k} \sum\limits_{i\in P_k} f_i(w).
\end{equation*}

The right-hand side of the above equation estimates the weight parameters for each smart doorbell device based on the loss values recorded across every data point (i.e., images) they trained with. The left side of the above equation scales each of those parameters and sums them all component-wise.


We have implemented the Federated Server using the Flask framework~\cite{flaskcomm} and hosted it on Amazon EC2. The Flask framework comes with an inbuilt web server. However, it is a single-threaded server, which is not ideal for our scenarios as the Federated Server has to handle multiple requests from Federated Clients. Therefore, we containerize the FlaskApp with Nginx~\cite{nginx} and Gunicorn~\cite{gunicorn}. The Gunicorn can handle multiple requests simultaneously. As a developer, you can configure Gunicorn with a number of workers and a number of threads it can run. These two parameters determine how much transactions you can handle at one point of time. The objective of using Nginx is to isolate the Federated Server logic from the Federated Clients. Second, it can act as a load balancer. Moreover, it can buffer multiple requests from clients and pass them to Gunicorn for further processing. This web application receives local model parameters from each client using HTTP POST request and distributes the global aggregated model back to each client. 



\subsection{Serverless Architecture}\label{sec:cloud}
One of our design goals is to minimize video data transmissions to the Cloud to reduce cost. However, we still need to store important video data to access data remotely. Therefore, we use the cloud to store detection results. For the sake of completeness, we briefly present the functionality of a doorbell hosted on the serverless infrastructure of Cloud. For the detailed description, we recommend the readers to refer our work:


\fakeparagraph{Real-time Push Notification} It sends a real-time alert notification to the user when a motion is detected in the proximity of the doorbell. We implement AWS Lambda functions that process the metadata in response to data ingestion~\cite{GYRARD2017305} from Kinesis and triggers the push notification~(Circled~\circled{1}--\circled{3} in Figure~\ref{fig:fl-server-cloud}), which is implemented using Amazon Simple Notification Service.


\fakeparagraph{Persistent Data Storage and Access} It receives video analytics metadata from a doorbell and provides a scalable storage to access data anywhere and anytime. We implement the storage services using Amazon DynamoDB and Amazon S3~(Circled~\circled{4} in Figure~\ref{fig:fl-server-cloud}), which are  exposed by Amazon API Gateway~(Circled~\circled{5} in Figure~\ref{fig:fl-server-cloud}), which accommodates the requests from MobileApp.


\fakeparagraph{Conversational User Interface} The voice assistant system leverages the logged video analytics results to provide a meaningful response. We implement an Alexa skill that can be triggered using the various voice commands~(such as ``\textit{Alexa, tell me what is happening at the door?}'', ``\textit{Alexa, send me a snapshot of all activities at my door today}'').  Our custom Alexa skill triggers a set of lambda functions, which queries the video analytics metadata stored in DynamoDB~(Circled~\circled{6} in Figure~\ref{fig:fl-server-cloud}). Once the query result is computed, the results are sent back through Alexa Voice.

\section{Demonstration}\label{sec:demo}


At the conference, we plan to demonstrate the following use cases:

\fakeparagraph{Use case 1: End-to-End Federated Learning Process for Video Analytics}
It demonstrates an end-to-end Federated Learning process, implemented for the smart doorbell case study. It consists of transmitting the model parameters from each smart doorbell device after local model training. The updated model parameters are stored at the Federated Server as files. The federated Server combines these local model parameters and generates a global aggregated model, which is eventually distributed to each smart doorbell in the federation to be used for inference in object detection.

\begin{figure}[h]
  \centering
    \includegraphics[width=8cm, height=5cm]{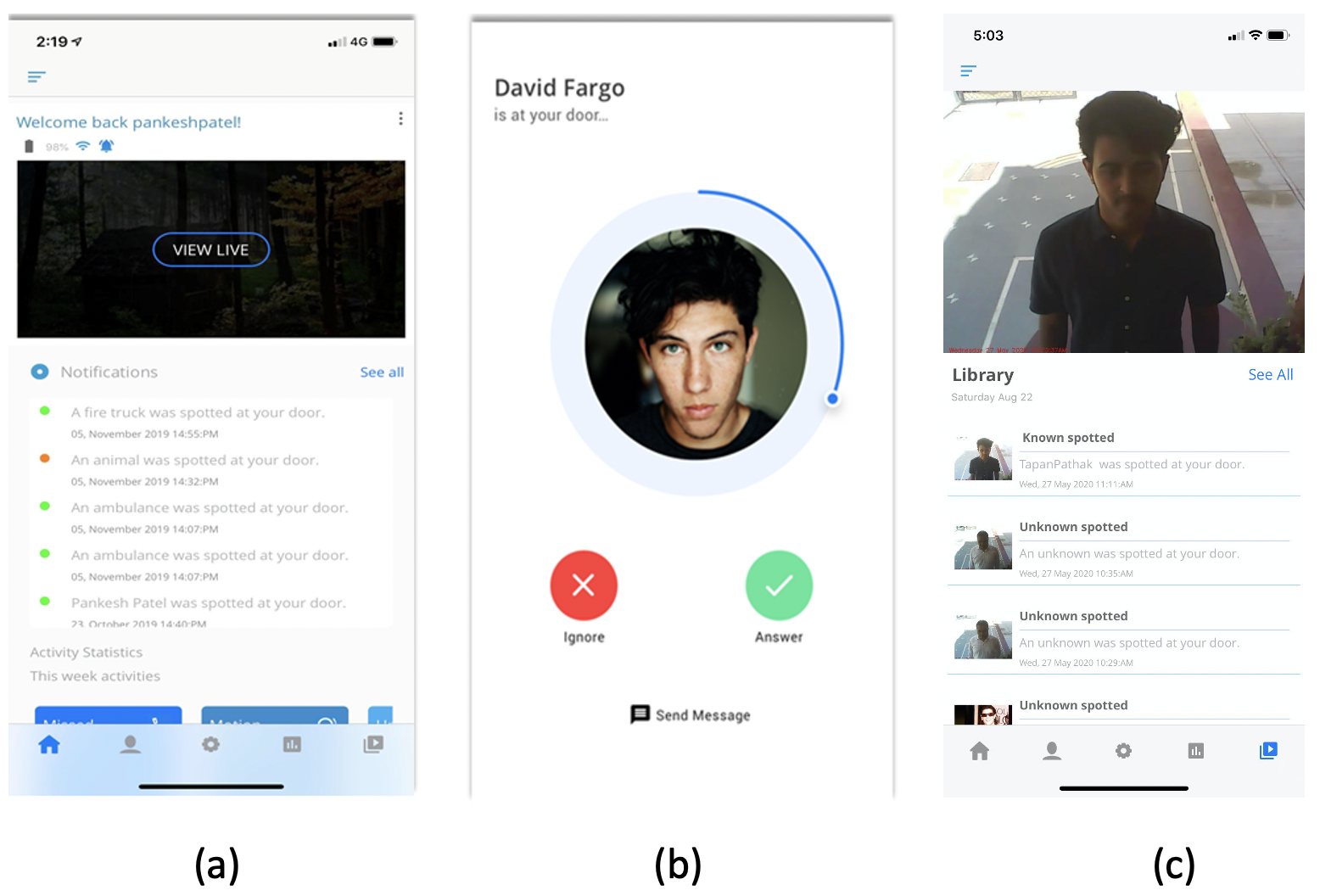}
  \caption{Smart Doorbell Mobile App.}
  \label{fig:mobileapp}
\end{figure}

\fakeparagraph{Use case 2: Object Detection using Global Federated Model} It demonstrates the live object detections by the doorbell. The system is initially at rest. An object entering the proximity of the doorbell enables the smart doorbell to start. This activity automatically triggers the object detection and recognition. The lower part of Figure~\ref{fig:mobileapp}--(a) shows the MobileApp dashboard that provides the detailed activities at the doorbell. The notification messages include face recognition (including known and unknown persons) and object detection (e.g., noteworthy car, animal, etc.). 

\fakeparagraph{Use case 3: Real-time Notifications using Global Federated Model} It demonstrates the ability of sending real-time alerts to the user when a motion is detected in the proximity of the doorbell. Figure~\ref{fig:mobileapp}--(b) shows an interface for real-time push notification. The user receives alerts on his mobile application when a visitor is detected at the door. The user can respond to the notification or just "ignore" it. Figure~\ref{fig:mobileapp}--(c) shows the video library. This interface of the MobileApp lets the users review activities and events at the door at a later time in case the user misses the real-time alert.

\fakeparagraph{Attendee Interactions} To demonstrate the Federated Learning based Smart doorbell design, we will carry three smart doorbell devices with us. The smart doorbell devices will be used to demonstrate the functionality of Federated Clients. Moreover, they will be used to present the smart doorbell hardware and software design and to explain how different components of the system interact with each other. Moreover, we will demo our work to explain the overall functionality of the doorbell. We will invite conference participants who are willing to try our MobileApp that lets them interact with the intelligent doorbell. We will keep a QR code at the booth to help install our MobileApp. To create an efficient flow of people at the time of demonstration, we will have a video played in loop on a laptop that we will bring along with us. 


\fakeparagraph{Technical Requirements} For demonstration at the conference, we will carry the required set of Raspberry Pi kit with sensors to demonstrate the FL-based smart doorbell functionality, an iPhone to interact with the smart doorbell, and a laptop to demo a web smart doorbell interface. From the conference organizers, we would only require a reliable WiFi/Ethernet internet to connect the smart doorbell to the software components running on AWS. 

 \section{Conclusion and Future Work}\label{sec:conclusion}
Through this paper, we demonstrate an intelligent smart doorbell design using Federated Learning across edge and cloud resources. The proposed smart doorbell design reduces communication cost, as smart doorbell uploads a trained model parameters to the centralized server, instead of images.  Second, the smart doorbell deploys On-Device Federated model~(aggregated by the Federated Server) to reduce the object detection latency.  Finally, it exchanges model instead of exchanging images, which provide with a sense of preserving privacy.

We understand that ``model aggregation rather than data aggregation'' is not enough to address the user’s privacy concerns fully. As a part of our future, we plan to extend the existing prototype on \textit{homomorphic encryption} where computing is done on encrypted image data, \textit{Secure Multiparty Computation~(SMC)} that enables multiple parties~(i.e., smart doorbell devices deployed across a large building) to collaboratively compute an agreed-upon computation without leaking information from participants. 





\begin{acks}
This publication has emanated from research supported by grants from the European Union’s Horizon 2020 research and innovation programme under grant agreement number 847577 (SMART 4.0 Marie Sklodowska-Curie actions COFUND) and from Science Foundation Ireland (SFI) under grant number SFI/16/RC/3918 (Confirm) cofunded by the European Regional Development Fund.
\end{acks}


\balance
\bibliographystyle{ACM-Reference-Format}
\bibliography{sample-base}

\end{document}